\documentstyle[prc,aps]{revtex}
\begin{document}
\draft
\title{ Microscopic mass formulae}
\author{J. Duflo$^a$ and A.P. Zuker$^b$}

\address{(a) Centre de Spectrom\'etrie Nucl\'eaire et de Spectrom\'etrie
de Masse (IN2P3-CNRS) 91405 Orsay Campus, France}

\address{(b) Physique Th\'eorique, B\^at40/1 CRN,
  IN2P3-CNRS/Universit\'e Louis Pasteur BP 28, F-67037 Strasbourg
  Cedex 2, France}

\date{\today}
\maketitle

\begin{abstract}
  By assuming the existence of a pseudopotential smooth enough to do
  Hartree-Fock variations and good enough to describe nuclear
  structure, we construct mass formulae that rely on general scaling
  arguments and on a schematic reading of shell model calculations.
  Fits to 1751 known binding energies for N,Z$\geq 8$ lead to rms
  errors of 375 keV with 28 parameters. Tests of the extrapolation
  properties are passed successfully.  The Bethe-Weizs\"acker formula
  is shown to be the asymptotic limit of the present one(s). The
  surface energy of nuclear matter turns out to be probably smaller
  than currently accepted.
\end{abstract}


The local structure of the mass surfaces \cite{audi} is fairly smooth
and amenable to algebraic analyses of shell model origin \cite{GK,LZ}
that lead to precise mass formulae - the rms errors may go below 400
keV \cite{JM} - but they need many parameters and have questionable
extrapolation properties far from stability.  The droplet (FRDM)
\cite{FRDM} and Thomas-Fermi (EFTSI) \cite{EFTSI} mass tables are
designed to reach the neutron drip-line, as demanded by calculations
of r-process nucleosynthesis, but they are not very precise: with 30
parameters the former yields an rms error of 673 keV for $ N,Z \geq 8$
while the latter needs only 9 parameters for an rms error of 730 keV
for $A\geq 36$ .

In a recent fit \cite{D} it was shown that with 12 parameters the rms
error for $N\geq 28$, $Z\geq 20$ could go down to 385 keV. This fit
does not extrapolate well and assumes a-priori knowledge of the right
closures and the right boundaries between spherical and deformed
regions. In a companion paper \cite{Z} rigorous microscopic guidelines
were proposed to take advantage of the simplicity detected in
\cite{D}, while eliminating the drawbacks. A brief review of these
guidelines will make clear which is the problem to be solved to obtain
a good mass formula.

The starting point is a basic separation property of the interaction
(anticipated in \cite{ACZ} but proven in \cite{Z} ): A (renormalized)
Hamiltonian, {\em ready for use in a shell model calculation} can be
written as $H=H_m+H_M$.

The monopole part,
\begin{equation}
   H_m=\sum_{k,l} a_{kl} m_{k}(m_{l}-{\delta}_{kl})
+b_{kl}(T_{k}.T_{l}-{3m_{k}\over4}{\delta}_{kl})
\end{equation}
contains only quadratic forms in the number ($m_k$) and isospin
($T_k$) operators for orbits $k$. The expectation value of $H_m$ for
any state is the average energy of the configuration to which it
belongs (a configuration is a set of states with fixed $m$ and $T$ for
each orbit).  In particular $H_m$ reproduces the exact energy of
closed shells and single particle (or hole) states built on them,
since for this set (which we call $cs\pm 1$) each configuration
contains a single member. We have emphasized the fact that eq.(1) is
ready for shell model calculations because its form is not the usual
one, in which effective core and single particle terms are separated
from the two body parts acting in the valence space. It should be
noted that the kinetic energy can always be written as two-body by
eliminating the center of mass contribution.

The multipole Hamiltonian $H_M$ contains all other terms (pairing,
quadrupole,etc.), and does not affect the $cs\pm 1$ states that bound
the shell model spaces.

This general result depends only on the assumption that there exists a
pseudo-potential smooth enough to do Hartree-Fock (HF) variation.  The
renormalization process that defines the effective interaction leads
also to many body forces that vanish at the closed shells.  Their
effects cannot be distinguished from those of the other higher rank
terms that will appear later in simulating configuration mixing
\cite{Z}.

In \cite{ACZ} it was shown that $H_M$ is well given by the realistic
nucleon-nucleon interactions, and therefore parameter-free while $H_m$
must be treated phenomenologically because of the poor saturation
properties of these forces.  The problem is that eq.(1) contains
hundreds of parameters, each of them a function of $A$ and $T$ ($T=|
N-Z |/2 $). In the context of \cite{ACZ} local fits are possible
because only a fraction of the parameters are active and information
on the $cs\pm 1$ states is available. To say something general about
$H_m$ and, in particular, to construct a mass table, we must proceed
differently.

In section I we shall find how to extract from eq.(1) the few
dominant terms responsible  for bulk properties and shell formation.
In section II we show how to mock configuration mixing through simple
algebraic expressions, and how to determine the spherical-deformed
boundaries. Section III contains the results of a fit, whose asymptotic
behaviour is discussed in section IV.

 {\bf I The form of $H_m$.}

To discover the  form of the dominant terms in $H_m$ we rely on
two geometric properties of the realistic forces.

{\sl 1.The $A,T$ dependence}. The $a_{kl}$ and $b_{kl}$ coefficients
 in eq.(1) must behave as typical matrix elements
\begin{equation}
 V(\omega)_{klmn}\cong {\omega\over{\omega}_0}
V({\omega}_0)_{klmn}+O({\omega}^{2}),
\end{equation}
a result from ref.\onlinecite{ACZ},but adding an $O({\omega}^{2})$
correction
warranted for large oscillator constant $\omega$. Then we know that
\begin{equation}
 \hbar\omega={34.6A^{1/3}\over <r^{2}>}\cong 40A^{-1/3}
\end{equation}
a classical result \cite{BM} obtained with the standard
 $<r^2>=0.86A^{2/3}.\quad$
A much better fit to the radii is obtained with $<r^2>=0.90{R_c}^2\quad$
where $R_c=A^{1/3}(1-(T/A)^2)$ \cite{D} .

By combining this result with eqs.(2,3), and calling a typical term in eq.(1)
 $\hat{\Gamma}\Gamma(A,T)$,  we obtain
\begin{equation}
 \Gamma(A,T)=(\Gamma/R)(1-\rho(\Gamma)/R))\quad,\quad R={R_c}^2/A^{1/3}
\end{equation}
for the amplitude $\Gamma(A,T)$ affecting operator $\hat{\Gamma}$.
(The hat $\hat{ }$ will be omitted when no  confusion is possible).

{\sl 2. Diagonal form. The isoscalar master terms}. The matrices $a_{kl}$
 and $b_{kl}$ in eq.(1) can be
reduced to diagonal form. For  $a_{kl}$, say,  we have
\begin{equation}
\sum_{k,l}a_{kl} m_{k}m_{l} = \sum_{\mu} e_{\mu}(\sum_{k}m_k
f_{k_\mu})^2
\end{equation}
and we borrow from \cite{DZ} the result that the largest contribution
(by far) to a realistic force has the form:
\begin{equation}
{\hat\Gamma_0}\Gamma_0(A)=(\sum_p m_p/\sqrt{D_p})^2 e_0,
\end{equation}
where $m_p$ is the number operator for the oscillator shell of principal
quantum number $p$ ,and $D_p$ the degeneracy

($D_p=(p+1)(p+2)$). Calling $M=\sum m_p/\sqrt{D_p}$,
setting $m_{p}=n_{p}+z_{p} $, where $n$, $z$ are number operators for
neutrons($\nu$) and protons ($\pi$), and filling shells $p_{\nu}$ and
$p_{\pi}$ up to some $p_{f\nu}$ and $p_{f\pi}$ Fermi level, we find :
\begin{equation}
<M>\cong{1\over2}[(3N)^{2/3}+(3Z)^{2/3}]=O(A^{2/3})
\end{equation}
where we have approximated $\sqrt{(p+1)(p+2)}\approx p+3/2 $, and used
 $N = \sum n_{p_\nu} = \sum (p_{\nu}+1)(p_{\nu}+2)\cong {1\over3}
(p_{f_{\nu}}+2)^3$ and $Z= {1\over3}(p_{f_\pi}+2)^3$.

Therefore ${<{\hat\Gamma_0}>}=O(A^{4/3})$, and, from eq.(4) we have
${<{\hat\Gamma_0}>\Gamma(A,T)}=O(A)$. As we shall see, this term
is responsible for most of the bulk energy. At the same time it
produces strong magicity at the harmonic oscillator (HO) closures
through the  $\sqrt{D_p}$ scaling of the $m_p$ operators.
This effect is absent in an infinite medium, where $a_{kl}=O(A^{-1})$
and the total number operator $m=\sum{m_p}$  replaces the eigenvector
 $M$ to ensure the $O(A)$ behaviour of the energy.

If $M$ is viewed as a coherent state, by symmetry the other
eigenvectors
 that can be constructed out of $m_p$ operators are simply these same
 operators
orthogonalized to $M$. Then we expect another monopole term of the form
$P=\sum f(p){m_p}^2/D_p$, which goes either as $A^{2/3}$ if $f(p)=O(1)$,
or as $A$ if $f(p)=O(p)$.

By now we can state the general rule:

{\bf Monopole terms must be symmetric quadratic forms of properly scaled
operators.}

Proper scaling is determined by the assumed form of the amplitudes (i.e.
eq.(4)), and by the asymptotic properties expected (or postulated) for
each term.
It is worth pointing out that proper scaling is a delicate matter that must
be decided on empirical grounds, and to illustrate the possible ambiguities
we consider a particular $a_{kl}$  coefficient in eq.(1) and take
$k$ and $l$ to belong to the same major shell.
 As contributor to the main
term $M^2$ (i.e. eq.(6)) $a_{kl}$ must contain a part that goes
 as $\omega/D_p=O(A^{-1})$, but it also contains
 a contribution  to  $P=\sum f(p){m_p}^2/D_p$, that goes
as $f(p)\omega/D_p$, i.e. $O(A^{-1})$  if $f(p)=O(1)$,
or  $O(A^{-2/3})$ if $f(p)=O(p)$. (In ref. \cite{DA83},  an $A^{-0.75}$
scaling is extracted from experimental spectra, quite consistent with
$f(p)=O(p)$).

\smallskip

{\sl Spin-Orbit and isovector terms}. In deciding which are the
 contributions
that are necessarily present beyond the ones already discovered,
 we consider
eigenvectors orthogonal to $M$. To obtain the observed extruder-
intruder closures (EI) we need spin-orbit (SO) effects and we introduce
\begin{equation}
S_p=pm_{jp}-2m_{rp}=(\widetilde{l\cdot s})_p.
\end{equation}
Here (refer to Fig.1) $jp$ is the largest orbit in the $p$-th shell
and $rp$ regroups all the others. $(\widetilde{l\cdot s})_p$ is the
operator
 that produces the same splittings as $(l\cdot s)_p$ and then collapses
 the
$r$-orbits to their centroid value. The rationale for considering only
two types of orbits is clear from Fig.1 : we want to give top priority to
shell formation (i.e. the $cs$ part of the $cs \pm 1$ set ).The combinations
of $m_k$ operators other than $m_j$ and $m_r$ will contribute to subshell
 effects that we incorporate in $H_M$ and treat later. It should be noted
 the formation of the EI closures is not inconsistent with the survival of
 HO ones in the ligth nuclei.

The other necessary ingredients are the isovector counterparts of $m_p$ and
$s_p$.
Calling $ \quad t_p=\vert n_p-z_p\vert$,

$S_p=p(n_{jp}+z_{jp})-2(n_{rp}+z_{rp})$ , $\quad$ and

$St_p=p\vert n_{jp}-z_{jp}\vert -2\vert n_{rp}-z_{rp}\vert$ ,

we introduce the variables
\begin{mathletters}
\begin{equation}
  MA_p ={ m_p \over\sqrt{D_p}}\quad,\quad SA_p={S_p\over 2(p+1)}\quad,
\end{equation}
\begin{equation}
  MT_p ={ t_p \over\sqrt{D_p}}\quad,\quad ST_p={St_p\over 2(p+1)}\quad.
\end{equation}
\end{mathletters}
 Finally we construct the 12 possible symmetric quadratics shown in
the first part of table I. The abreviations are ($F$=full,$P$=partial),
($M$=master, $S$=SO, $C$=cross), ($A$=isoscalar, $T$=isovector).
The $C$ terms  are hybrids
 that could not represent eigensolutions in eq.(5), but they could
modify (i.e. mix) slightly the $M$ and $S$ terms. They have been introduced
to ensure - through $FCA$ and $FCT$ - the presence of representatives
 of the conventional spin-orbit force, including possible isospin effects.
 It is only for ``symmetry '' reasons in table I that $PCA$ and $PCT$ are also
 present, but their effects are not expected to be significant and they will
 excluded from the fits considered in this paper.

 $FSA$ and $FCA$ are scaled so to go as the ordinary
 $\zeta\, l\cdot s$ term with $\zeta = O(A^{-2/3})$ \cite{BM}. The $FXT$
 operators are scaled as the $FXA$ ones. For the $PXY$ contributions
we introduce a factor $D^{\alpha}_{p}$ to allow for  possible ambiguities
in the $f(p)$ factor discussed above.
 The fits will decide in favour of $\alpha=1/2$.

In addition to these terms, $H_m$ includes a pairing ($V_p$) and
Coulomb ($V_c$) contributions as well as a $4T(T+1)$ term whose presence
we explain below. Although $V_p$ belongs in principle with $H_M$, it
has been incorporated here because it affects equally all nuclei. Similarly
$V_c$ contains multipole parts, but at the level of precission we work they
are negligible.

{\sl Asymtotic form of the master terms}.
Except for a small contribution  from $PMA$ and $PMT$ (when $\alpha =
1/2$), the asymptotic expressions

$\quad FMA\asymp{({3\over2} A)}^{4/3}(1-{2\over9}{(2T/A)}^2) \quad $ and

$\quad FMT\asymp{({3\over2} A)}^{4/3}{({4T\over{3A}})}^2 $

determine
the bulk and  symmetry energies of nuclear matter. As these
quantities must emerge from a balance of potential and kinetic
energies that our  arguments
%onecolumn
%begin{minipage}[t]{18.0 cm}
\begin{table}
\caption{The operators $\hat\Gamma$ ({\it called $\Gamma$ here})
in $H_m$, $H_s$ and $H_d$ }
\begin{tabular}{lll}
\tableline
 $H_m$&$\alpha=1/2$&$R_c=A^{1/3}(1-
 (T/A)^2) $\\
 $FMA=(\sum MA_p)^2$&$FSA=(\sum SA_p)^2$&$FCA=\sum MA_pD_p^{-1/2}\sum
 SA_pD_p^{-1/2}$\\
 $PMA=\sum (MA_p)^2D_{p}^{\alpha}$&$PSA=\sum (SA_p)^2D_{p}^{\alpha}$&
 $PCA=\sum (MA_p)(SA_p)D_p^{\alpha -1}$\\
 $FMT=(\sum MT_p)^2$&$FST=(\sum ST_p)^2$&
 $FCT=\sum MT_pD_p^{-1/2}\sum ST_pD_p^{-1/2}$\\
 $PMT=\sum (MT_p)^2D_{p}^{\alpha}$&$PST=\sum (ST_p)^2D_{p}^{\alpha}$&
 $PCT=\sum (MT_p)(ST_p)D_p^{\alpha -1}$\\
 $4T(T+1)A^{-2/3}$&$V_p=-mod(N,2)-mod(Z,2)$&
 $V_c=[-Z(Z-1)+.76[Z(Z-1)]^{2/3}]/R_c$ \\
 \------------&&\\
 $H_s \quad(\bar{n}=D_{\nu}-n,\quad \bar{z}=D_{\pi}-z)$&
 $S2=n\bar{n}D_{\nu}^{-1}+z\bar{z}D_{\pi}^{-1}$&
 $S3=n\bar{n}(n-\bar{n})D_{\nu}^{2(\beta -1)}+z\bar{z}(z-\bar{z})
 D_{\pi}^{2(\beta -1)}$\\
 $SQ+=2(n\bar{n})^2D_{\nu}^{2\beta -3}+2(z\bar{z})^2D_{\pi}^{2\beta -3}$&
$SQ-=4n\bar{n}z\bar{z}(D_{\nu}D_{\pi})^{\beta -3/2}$&$\beta=1/2$\\
\------------&&\\
 $H_d \quad (n'=n-JU , \bar{n}'=\bar{n}+JU) ,$&
 $(z'=z-JU, \bar{z}'=\bar{z}+JU)$&
 $D3=n'\bar{n}'(n'-\bar{n}')D_{\nu}^{-2}+z'\bar{z'}(z'-\bar{z}')
 D_{\pi}^{-2}$\\
 $QQ_1^0=(n'\bar{n}'D_{\nu}^{-3/2}\pm z'\bar{z}'D_{\pi}^{-3/2})^2$&
 $QQ\pm =QQ0\pm QQ1$ &$D0=16, JU=4.$\\
\end{tabular}
\end{table}
%end{minipage}
%begin{minipage}[t]{8.6 cm}
 have bypassed, the shell effects produced by $FMA$ and  $FMT$
may be partially spurious and it is convenient to decouple them from
the asymptotic contributions by adding these as independent terms.

 The three operators that turn out to be favoured by the fits are
  $4T(T+1)/A^{2/3}$, and $FM+$ and $PM+$
where $XM+=XMA+XMT,\quad (X\equiv F,P),\quad$ whose asymptotic form is
\begin{equation}
 XM+ =1.717A^{4/3}+.3816(2T)^2/A^{2/3} +O(A^{2/3})
\end{equation}

{\bf II Configuration mixing due to $H_M$ }

Once $H_m$ has ensured shell formation, mostly of EI type,
 the variables that become
important in  modelling configuration mixing are $n_v$ and
$z_v$, the number of valence particles in EI spaces of degeneracy
 $D_\nu$ and $D_\pi$ (see Fig.1). For the light nuclei these model spaces are
 unconventional but not necessarily wrong: $^{12}C$ can be a good core
 \cite{ZBM}, and the $N,Z$=14 closures are at least as convincing as the
 $N,Z$=20 ones \cite{Z}.

\smallskip

{\sl Spherical nuclei.$H_s$and subshell effects}.
The energies of the spherical nuclei will be mocked by $H_s$, a linear
combination of the four operators  $S2, S3, SQ+, SQ-$,
listed in the second part of table I,
  each affected by a coefficient as given in
eq.(4). As explained in \cite{Z}, $H_s$ is largely devoted to smooth
subshell effects. Since $S2$ is easily absorbed in $H_m$, and $SQ+$ is
not expected (or found) to be important, we are left with only two terms
 whose main task is to correct the artificial filling patterns
due to the imposed degeneracy of the $r$-orbits
(we shall return to this point in section III and in the conclusion).
The $D^{\beta}$  factors in table I reflect scaling uncertainties
associated to these corrections.

\smallskip

{\sl Deformed nuclei.$H_d$}. The energies of permanently
deformed ground states
will be described by $H_d$ whose form was not derived in \cite{Z}, but it can
be easily shown that it must must be similar to that of $H_s$, although
the underlying physics is different. The argument is that the onset
of rotational motion
is associated with the interruption of normal spherical filling by the
promotion of $JU_{\pi}$ protons and $JU_{\nu}$ neutrons to
 configurations of the next HO shell.  Nilsson diagrams
indicate that  $JU_{\pi}$ =$JU_{\nu}$=$JU$=4. The intruders bring a loss
 of monopole and a gain of quadupole energy represented by the constant
 $D0$, (the number 16 in table I is an arbitrary factor).
The deformation energy of the particles in the
lower orbits is simulated by quadrupole terms $QQ+\text{ and }QQ-$,
corresponding to
the filling of equidistant Nilsson orbits, and $D3$ is in charge of the
balance of monopole effects. Scaling is such that deformation energies
have the standard $A^{1/3}$ behaviour.

\smallskip

{\sl The fitting procedure and the spherical-deformed boundaries}.
Energies taken to be positive are given by the expectation values
\[ E(N,Z)=<H_m>+<H_s>(1-\delta_d)+<H_d>\delta_d\]
\begin{equation}
 =max(<H_m>+<H_s> , <H_m>+<H_d>)
\end{equation}
%end{minipage}
%begin{minipage}[t]{18.0 cm}
\begin{table}
\caption{Parameters of the $28p$ and $28p^\star$ fits / $\cal A$
 $\Rightarrow FM+$ $(28p)$
 or $A^{4/3}$ $(28p^\star)$ / $V_p$ and $V_c$ given in Table 3.
 The numerical values correspond to binding energies in MeV.}
\begin{tabular}{ldddddddddddddd}
$\hat\Gamma$&$\cal A$&PM+&4T(T+1)&FS+&FC+&S3&D0&QQ-&PS+&PS-&FS-&D3&
SQ-&QQ+\\
\tableline
 $\Gamma 28$&9.55&-0.77&-37.23&6.03&-11.18&0.47&-38.1&25.5
  &-0.9&-0.13&1.4&-0.9&0.35&4.6\\
 $\rho(\Gamma)28$&0.89&1.3&1.38&4.55&5.11&4.75&4.81&4.09
  &5.24&5.03&4.21&0.&4.47&0.\\
 $\Gamma 28^\star$&16.73&-0.78&-33.35&6.05&-18.04&0.42&-39.8&13.6
  &-1.2&-0.16&1.7&0.3&0.16&3.3\\
 $\rho(\Gamma)28^\star$&1.44&4.87&1.45&5.39&4.13&4.34&4.79&2.75
  &6.09&5.30&4.24&0.&4.35&0.\\
\end{tabular}
\end{table}
%end{minipage}
%begin{minipage}[t]{8.6 cm}
the lowest possible orbits are filled for spherical nuclei
($\delta d=0$),
 while for deformed ones ($\delta d=1$), $JU$ particles are promoted to
orbits $j$. The calculations are conducted
by initializing $\delta d$, fitting $E(N,Z)$ in the first
equality of eq.(11) to the 1751 mass values for $N,Z\geq 8$ in the latest
compilation \cite{audi}, then resetting $\delta d$ through
the second part of eq.(11) and iterating until convergence.
Deformed states are only allowed in the regions where neutrons and
 protons
fill different orbits i.e. above $A\approx100$. Below, the onset of
rotational motion is not sharp, pairing effects are stronger and a
 specific
treatment would be called for.

{\bf III The fit.}
It has already been mentioned that only a particular combination of the
master terms was favoured, and that the $S2$, $SQ+$, $PCA$ and $PCT$
contributions would be omitted. In addition, we keep only one $XCY$ term
and omit the $\rho$ factors in $D3$ and $QQ+$ (the less significant of all
the included operators). This leaves us with 28 parameters, shown in table II.
This fit, $28p$, yields an rms error of 375 keV.
In the $28p^\star$ fit,  $\widehat{FM+}$ is replaced
by $A^{4/3}$, leading to an rms error of 485 keV. The interest of
 comparing  $28p$ and $28p^\star$ will become clear in section IV.

  The notation
for the operators is $XY\pm =XYA\pm XYT$ and we have preset $JU=4$,
 $\alpha=\beta=1/2$.

It is clear that the $\rho$ ratios fall in two broad categories: for
the master terms they are close to unity, and for all others they bunch
in the interval 4.5$\pm$1. According to
eq.(4), it means that the corresponding contributions change sign at
$\rho\approx A^{1/3}$, i.e. $A\approx100$,
 {\em the region where the $j$ orbits can start filling
 before the $r$ ones are full}. This is a clear indication that
 parametrization (4) acts in a way unrelated to the derivation in section I,
and that all contributions - in addition to doing what they were designed for -
are busy simulating subshell effects.
Setting $\rho$=0 is in general of little consequence , except for $S3$, which
entails a loss of 300 keV, and to a lesser extent $QQ-$ (loss of 150 keV); and
it is possible to obtain fits with rms errors of some 600 keV, with only a
dozen parameters.

 Some elements of evaluation are given next.

{\it Radii}.
 The radius extracted from $V_c$, $r\approx1.235R_c$ is close
to the fitted $r\approx1.225R_c$ \cite{D}. The use of $R_c$ rather than
 $A^{1/3}$ leads to a gain of 34 keV.

{\it Deformed Nuclei}. The number of deformed nuclei is 396
for $28p$ and  393 for
$28p^\star$, with rms errors of 254 and 313 keV respectively. The sharp
transitions (e.g. at $N=89,90$) are accurately reproduced.

{\it Extrapolation properties} When fitted to the 1503 nuclei in the
1983
compilation, $28p$ yields the rms error of 357 keV. When the same
parameter  are
used on the 248 new nuclei in the 1993 tables the rms error is 470 keV,
which amounts to good predictive power by present standards.
The increase in error is due only to spherical nuclei, and we have
found that discrepancies exceeding 700 keV are almost
invariably associated with subshell effects that are important
only in spherical regions. Progress in this matter can
be expected  through a more sophisticated treatment
of the $S$ operators.

When it comes to systematic behaviour, which
includes many spherical nuclei and all the deformed ones
(unsensitive to subshell details),   the fits do very well
 as emphasized by another test: when 28$p$ in table II is
used on the 2542 entries (i.e. including systematic trends)
of the 1993 tables the rms error is almost unchanged (407 keV).

{\bf IV Asymptotics and the Liquid-drop}.

 The $28p$ and $28p^{\star}$  entries in Table III are the asymptotic
contributions, given by eq.(10) for $FM+$ and $PM+$, plus the Coulomb
and pairing terms. Under $6p$ we have the results of a direct fit
to the data by varying these 6
%end{minipage} \ . \
%begin{minipage}[t]{8.6 cm}
\begin{table}
\caption{Asymptotic forms of the fits compared with a pure LD form
($6p$). (T2 is for 4T(T+1))}
\begin{tabular}{ldddddd}
 &$A^{4/3}/R$&$A^{4/3}/R^2$&$T2/A^{2/3}R$&$T2/A^{2/3}R^2$&$V_p$
 &$V_c$\\
\tableline
 $28p$&15.07&-12.79&-33.88&48.51&5.18&.699\\
 $28p^\star$&15.39&-17.54&-33.64&49.81&5.15&.699\\
  $6p$&15.42&-17.56&-33.65&50.91&5.18&.699\\
\end{tabular}
\end{table}
 parameters, that can be identified to the classical liquid drop
(LD) ones by expanding  the $R$ and $R^2$  denominators (rms error
 of the $6p$ fit: 2499 keV).

 The agreement between $28p$, $28p^{\star}$ and $6p$ in Table III is
 excellent - even spectacular - except for  the surface coefficient
  in $28p$.(Replacing $FM+$ by $FMA$ leads to an intermediate situation
with an rms error of 436 keV and a surface coefficient of -14.6).

 It is apparent that conventional LD parameters as extracted  in $6p$
 are fully reproduced by $28p^{\star}$, a very good fit by present
 standards. However, $28p$ is significantly better and provides a strong
 hint that the surface energy may be smaller than the conventional
 value.

    {\bf Conclusions and prospects. }

The rms errors we have obtained may seem good - even impressive - by
present standards, but they are still large compared to those obtained
in shell model calculations, i.e. some 300 keV in the lightest nuclei
and 150 keV around $A=100$ \cite{ACZ}. The  reason is that our
 treatment of $H_m$ is too crude since we have forced an artificial
 degeneracy on the $r$-orbits. As explained in ref.\cite{Z}, the correct
  behaviour can be simulated to some extent by the $H_s$ and $H_d$ terms
 - designed to describe configuration mixing - but also capable of coping with
  ``monopole drift'', i.e. the smooth part of subshell effects.

A more refined  treatment of the $S_p$ operators in $H_m$ will eliminate
the large $\rho$ ratios, which are physical - in the sense that they represent
a real effect- but mask the true contribution of a given term, and introduce
unnecessary phenomenology in the formulation.

It should be clear from these remarks that the fit(s) we have proposed
are only exploratory. There are grounds to expect much better ones.

{\it NOTE.} It takes only seconds to generate a mass table. The program
is available from dufloj@frcpn11.in2p3.fr .

\vspace{0.3 cm}
\acknowledgments
We would like to thank G. Audi, G.E. Brown, E. Caurier,
T.von Egidy, B. Jonson, P. M\"oller, H. Niefnecker, M. Pearson, A. Poves,
P. Quentin, A. Sobicewsky, F.K. Thielemann, F. Tondeur and N. Zeldes for
useful exchanges.
%end{minipage}

%onecolumn
%begin{minipage}[t]{18.0 cm}

%end{minipage}

\begin{figure}
\setlength\unitlength{0.5mm}
\begin{picture}(90,90)
\protect
\put(55,90){HO}
\protect
\put(90,90){EI}
\protect
\put(28,75){$D_{p+1}$}
\protect
\put(50,77){$p+1$}
\protect
\put(85,82){$r(p+1)$}
\protect
\put(45,75){\line(1,0){20}}
\protect
\put(65,75){\line(2,1){10}}
\protect
\put(65,75){\line(2,-3){10}}
\protect
\put(75,80){\line(1,0){35}}
\protect
\put(85,62){$j(p+1)$}
\protect
\put(75,60){\line(1,0){35}}
\protect
\put(112,53){$D_{\nu ,\pi}$}
\protect
\put(85,52){$r(p)$}
\protect
\put(32,42.5){$D_p$}
\protect
\put(50,44.5){$p$}
\protect
\put(40,42.5){\line(1,0){25}}
\protect
\put(65,42.5){\line(4,3){10}}
\protect
\put(65,42.5){\line(4,-5){10}}
\protect
\put(75,50){\line(1,0){35}}
\protect
\put(85,32){$j(p)$}
\protect
\put(75,30){\line(1,0){35}}
\protect
\put(0,20){$D_{jp}=2(p+1)\quad \quad D_{rp}=p(p+1)$}
\put(0,10){$ D_{\nu,\pi}=D_p+2 =D_v
\quad\quad n_{\nu,\pi}=n_v=n_{j(p+1)}+n_{rp}$}
\protect
\caption{ HO and EI major shells.}
\protect
\end{picture}
\end{figure}
\end{document}